\documentclass[twocolumn,trackchanges]{aastex7}
\usepackage{algorithm2e}
\SetKwProg{Fn}{Function}{:}{End Function}
\usepackage{orcidlink}
\usepackage{amssymb}
\usepackage{newtxtext,newtxmath}
\usepackage{siunitx}
\usepackage[utf8]{inputenc}
\usepackage[T1]{fontenc}
\usepackage{CJKutf8}
\usepackage{booktabs}

\usepackage{graphicx}	
\usepackage{amsmath}	

\SetKwComment{Comment}{/* }{ */}

\begin{document}

\title{The Universe Favors Primes: A Study in the Primality of Cosmic Structures}

\author[gname=Li N.]{Nan Li~\begin{CJK}{UTF8}{gbsn}(李楠)\end{CJK}
~\orcidlink{0000-0001-6800-7389}~\thanks{Email: \href{nan.li@nao.cas.cn}{nan.li@nao.cas.cn}}}
\affiliation{National Astronomical Observatories, Chinese Academy of Sciences, Beijing, 100101 China}
\email{nan.li@nao.cas.cn}~\thanks{Email: \href{nan.li@nao.cas.cn}{nan.li@nao.cas.cn}}

\author[gname=Shen S.]{Shiyin Shen~\begin{CJK}{UTF8}{gbsn}(沈世银)\end{CJK}
~\orcidlink{0000-0002-3073-5871}~\thanks{Email: \href{mailto:ssy@shao.ac.cn}{ssy@shao.ac.cn}}}
\affiliation{Shanghai Astronomical Observatory,  Chinese Academy of Sciences, Shanghai, 200030, China}
\email{ssy@shao.ac.cn}~\thanks{Email: \href{mailto:ssy@shao.ac.cn}{ssy@shao.ac.cn}}

\begin{abstract}
The cosmological principle states that the universe is uniform and does not favor any specific position or direction. However, research conducted by \cite{Shen2025} has revealed that the universe demonstrates a notable inclination towards parity-odd states. Furthermore, it remains uncertain whether the universe also favors prime numbers. In this study, we examine the largest available catalogs of galaxy groups to investigate this hypothesis. Specifically, we assess whether the number of galaxies within a galaxy group or cluster is more likely to be a prime number. Our results strongly suggest that the universe does indeed have a preference for prime numbers, with findings exceeding the 4.1 sigma significance threshold. This insight explains why the Primes consistently triumphs over Unicorn. Consequently, it may be necessary to consider revising the cosmological principle in the context of a higher-dimensional feature space. Moreover, our research establishes a connection between the Riemann Zeta function and cosmology pioneeringly, paving the way for the development of \textbf{Cosmozetaology}.
\end{abstract}
\keywords{Cosmology, Galaxy Group, Prime Number, Riemann Zeta Function, Transformers}

\section{Introduction}
The cosmological principle, also known as the Copernican Principle, is one of the foundational pillars of modern cosmology. It posits that the universe is homogeneous and isotropic on large scales, meaning there is no preferred location or direction. This principle has been instrumental in shaping our understanding of the universe's large-scale structure and its evolution. By assuming homogeneity and isotropy, astronomers have developed robust theoretical models, such as the Friedmann-Lemaître-Robertson-Walker (FLRW) metric, which describes the expansion of the universe. Observational data, such as those from large-scale structure surveys and cosmic microwave background missions, have also shown strong support for the homogeneity and isotropy of the universe\footnote{This paragraph is mainly written by AI.}. 

Nevertheless, the universe does not consistently exhibit randomness and symmetry throughout. For instance, several studies have shown that the distribution of spiral galaxies' spin directions lacks both randomness and symmetry. A greater proportion of galaxies in the northern hemisphere rotate counterclockwise, whereas the southern hemisphere exhibits the reverse trend\citep[][but also see \citet{Patel2024}]{Shamir22}. Besides, \citet{Shen2025} provides a robust analysis and concludes that the universe prefers odd numbers. Going deeper into the thoughts \citep[][]{Arefeva2007, Dittrich2019, Elizalde2021}, a more profound question arises: Does the universe favor prime numbers? 

According to the current standard model of cold dark matter cosmology, galaxies form within dark matter halos and grow through a process known as hierarchical merging. Since this merging process is non-linear and occurs over a long timescale, a dark matter halo typically contains more than one galaxy. In a random universe without inherent preferences, the number of galaxies within a galaxy group (or dark matter halo) should also be random. This means there should be no bias towards having a prime or composite number of member galaxies. Therefore, the probability of a galaxy group containing a prime number of member galaxies should align with the predictions made by the prime number distribution function $\pi(x)$. Is this hypothesis true? Is it possible that our universe has a preference for primality? This paper aims to provide a quick answer to this nontrivial question. 

The structure of this paper is as follows. Section \ref{section:data} introduces the data used in this work. In Section \ref{section:method}, we detail our method for calculating the parity of our universe using the datasets above and present our findings in Section \ref{section:result}. Lastly, our conclusions are drawn in the Section \ref{section:conclusion}. Throughout this study, we adopt the $\Lambda$CDM cosmology with the parameters specified by \cite{Planck2020}: $\Omega_{\Lambda}=0.6889$, $\Omega_{\rm m}=0.3111$, $\Omega_{\rm b}=0.04897$, $n=0.9665$, and Hubble constant $H_0 = 67.66$ km\,s$^{-1}$\,Mpc$^{-1}$.

\section{Data}
\label{section:data}
We leverage the group catalog derived from the DESI Legacy Imaging Surveys DR9, as referenced by \citet{Yang2021}. This catalog represents the most comprehensive collection of galaxy groups currently available, encompassing two sky regions: the south galactic cap (SGC) and the north galactic cap (NGC). Within these regions, we identify groups that contain at least three members, yielding a substantial sample of over 5.7 million groups and more than 30 million group members. The large size of this dataset facilitates a statistically robust analysis.

\section{method}
\label{section:method}

We start by counting the number of members in each group. Subsequently, we classify the groups into two categories: those with a prime number of members and those with a composite number of members. Once categorized, we calculate the fraction of prime numbers within three defined ranges: [10, 40), [40, 70), and [70, 100). Additionally, we derive theoretical predictions for the fractions of prime numbers using the Riemann zeta function R(x) over the same intervals. 
\begin{equation}
    R(x) = 1 + \sum_{k=1}^{\infty} \frac{(\ln x)^k}{k! \, k \, \zeta(k+1)}
\end{equation}
The algorithm's flow and the corresponding pseudo-code are detailed in Algorithm 1.

\RestyleAlgo{ruled}
\begin{algorithm}
\caption{Prime counting utilities}
\DontPrintSemicolon

\SetKwFunction{IsPrime}{is\_prime}
\SetKwFunction{Fraction}{fraction\_of\_primes\_in\_range}

\KwIn{An integer $n$}
\KwOut{True if $n$ is prime, False otherwise}
\Fn{\IsPrime{$n$}}{
    $n \leftarrow \text{int}(n)$\;
    \If{$n \le 1$}{
        \Return \textbf{false}\;
    }
    $limit \leftarrow \lfloor \sqrt{n} \rfloor$\;
    \For{$i \leftarrow 2$ \KwTo $limit$}{
        \If{$n \bmod i = 0$}{
            \Return \textbf{false}\;
        }
    }
    \Return \textbf{true}\;
}

\BlankLine
\BlankLine
\BlankLine

\KwIn{Integers $n_{\text{min}}$ and $n_{\text{max}}$ with $n_{\text{min}} < n_{\text{max}}$}
\KwOut{Fraction of primes in the range $[n_{\text{min}}, n_{\text{max}}]$}
\Fn{\Fraction{$n_{\text{min}}$, $n_{\text{max}}$}}{
    \If{$n_{\text{min}} \ge n_{\text{max}}$}{
        \Return $0.0$\;
    }
    $primeCount \leftarrow 0$\;
    \For{$n \leftarrow n_{\text{min}}$ \KwTo $n_{\text{max}}$}{
        \If{\IsPrime{$n$}}{
            $primeCount \leftarrow primeCount + 1$\;
        }
    }
    $totalCount \leftarrow n_{\text{max}} - n_{\text{min}} + 1$\;
    \Return $primeCount / totalCount$\;
}
\end{algorithm}

\begin{figure}[htb!]
\centering
\includegraphics[width=\columnwidth]{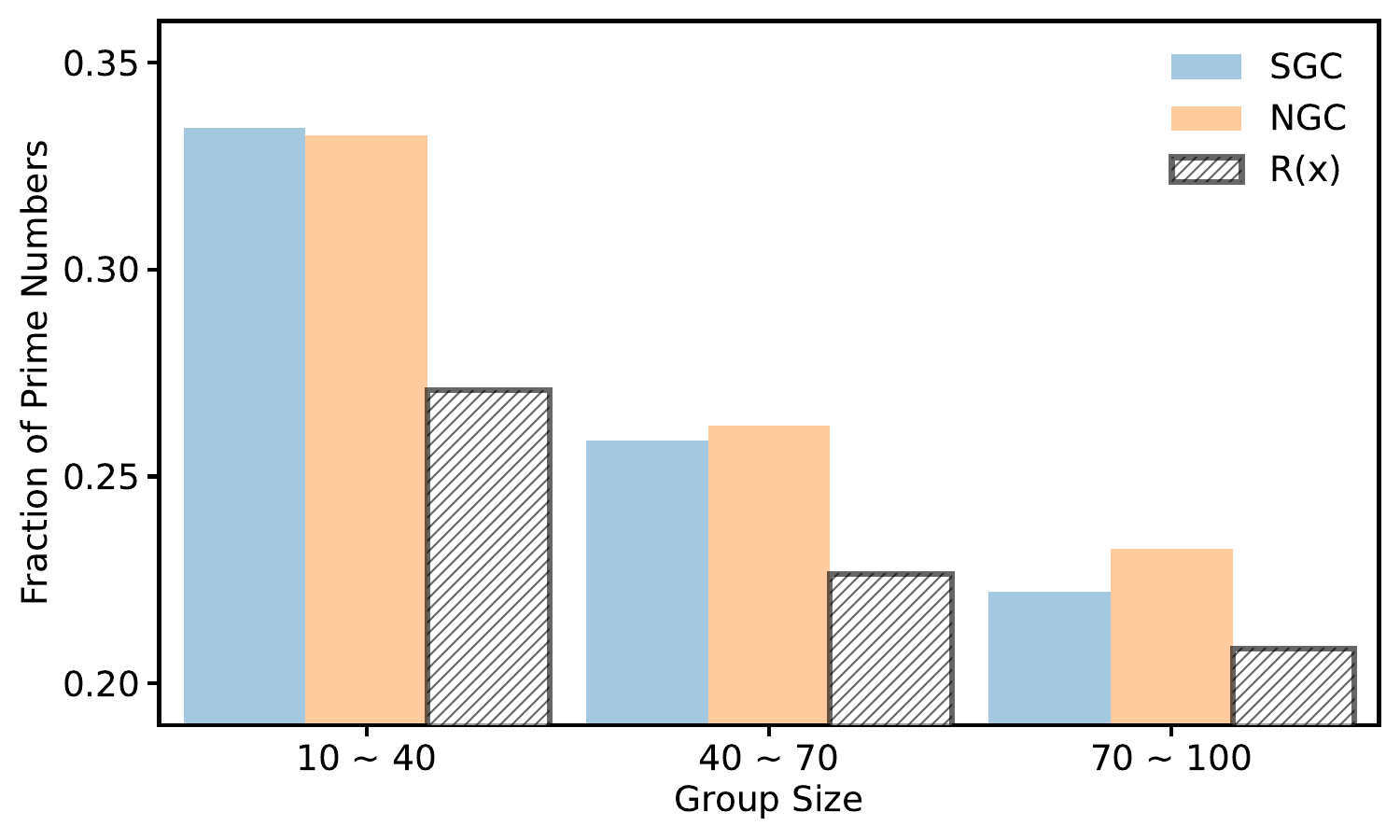}
    \caption{The number of DESI DR9 groups with prime and composite group members in the south galactic cap (SGC, blue bars) and the north galactic cap (NGC, orange bars) regions in different bins, respectively. Black hatched bars present the fractions of primes in the corresponding number bins given by the Riemann function $R(x)$.}
    \label{fig: fraction_of_primes}
\end{figure}

\section{Results}
\label{section:result}
For DESI DR9 groups with member counts in the ranges of [10, 40), [40, 70), and [70, 100), there are \{75156, 2290, 414\} and \{74931, 2333, 454\} groups located in the SGC and NGC area, respectively. The fractions of groups possessing a prime number of members are approximately 33\%, 26\%, and 22\% in both the SGC and NGC areas. However, the fractions predicted by the Riemann zeta function are 27\%, 23\%, and 20\% for these ranges. Figure 1 presents these findings in a bar plot. Notably, in both regions, the observed fractions of groups with a prime number of members consistently exceed the theoretical predictions, indicating a significant discrepancy that decreases with group size. When analyzed using simple Poisson statistics, this discrepancy is found to be significant at a level greater than $4.1 \sigma$.

\section{Conclusions}
\label{section:conclusion}

After conducting a comprehensive investigation into the distribution of prime numbers across various group sizes, we have uncovered a significant insight: the Universe appears to favor primes. This finding helps clarify why the Primes consistently defeated Unicorn, even when Primus is merely the remnants of a dying planet. As a result, it may be necessary to reconsider the cosmological principle within the framework of a higher-dimensional feature space. Moreover, our research forges a novel connection between number theory and cosmology pioneeringly, potentially paving the way for the development of \textbf{Cosmozetaology}.

\section*{Data Availability}
\label{sec: public}
 The code of this study is available at \url{https://github.com/linan7788626/the-universe-favors-primes} and the DESI DR9 group catalog is available at \url{https://gax.sjtu.edu.cn/data/data1/DESI_DR9/DESIDR9_member.tar.gz}.

\begin{acknowledgments}
This study is not supported by any scientific grants. SS made no academic contribution to this study and is included here solely to ensure the number of authors to be \textbf{Prime}. 
\end{acknowledgments}

\bibliographystyle{aasjournalv7}
\bibliography{sample7}

\end{document}